\documentclass[twocolumn,twoside,slac_two]{revtex4}
\usepackage{graphicx,wasysym}
\usepackage{fancyhdr}
\begin{document}

\title{The LHCb Upgrade}

%

\author{H.~Dijkstra\footnote{for the LHCb collaboration}}
\affiliation{CERN, CH-1211, Geneva 23, Switzerland}

\begin{abstract}
The LHCb detector has been designed to study CP violation and
other rare phenomena in B-meson decays up to a 
luminosity of $\sim 5.10^{32}\rm cm^{-2}s^{-1}$. This paper will describe
what is limiting LHCb to exploit the much higher luminosities available
at the LHC, and what are the baseline modifications which will remedy these
limitations. The aim of SuperLHCb is to increase the yields in hadronic 
B-decay channels by about a factor twenty compared to LHCb, while for
channels with leptons in the final state a factor ten increase in statistics 
is envisaged.
\end{abstract}

\maketitle

\thispagestyle{fancy}


\section{Introduction}
\begin{figure*}[t]
\centering
\includegraphics[width=135mm]{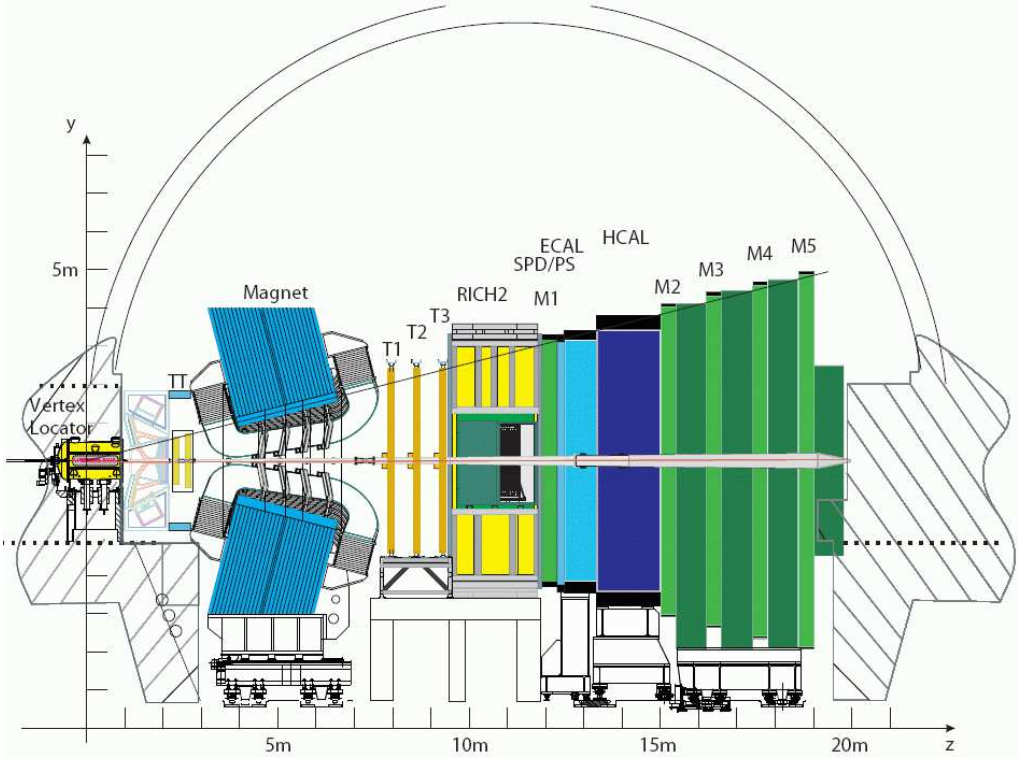}
\caption{LHCb detector layout, showing the Vertex Locator (VELO), 
the dipole magnet, the two RICH detectors, the four tracking stations 
TT, T1-T3, the Scintillating Pad Detector (SPD), Preshower (PS), Electromagnetic (ECAL) and Hadronic (HCAL) calorimeters, and the five muon stations M1-M5.} \label{LHCb}
\end{figure*}

The LHCb experiment has been conceived to study CP violation and other rare 
phenomena in the B-meson decays with very high precision. The experiment
is at the moment in the last phase of its construction, and is expected
to be fully operational when the LHC machine will deliver its
first pp collisions at 14 TeV in the summer of 2008. Figure~\ref{LHCb}
shows the layout of the detector, of which a detailed description can be 
found in ~\cite{reop}. The detector has been designed to be able to cope with
an instantaneous luminosity up to $\sim 5.10^{32}\rm cm^{-2}s^{-1}$, and
a total radiation dose corresponding to $\sim 20\rm~fb^{-1}$. After an initial
shake down of the detector in 2008, the aim is to look for New Physics (NP)
signatures compatible with luminosities around $\sim 0.5 \rm~fb^{-1}$.
The next four to five years, LHCb will accumulate $\sim 10\rm~fb^{-1}$ 
to exploit the full physics program envisaged for the present detector.
In the next section a selection will be presented of the expected performance
of LHCb within the aforementioned luminosity range.

As mentioned above, LHCb will run at luminosities a factor 20-50 below the
$10^{34}\rm cm^{-2}s^{-1}$ design luminosity of the LHC. The machine optics of LHCb do allow
to focus the beams sufficiently to run at luminosities a factor ten larger.
Hence, the upgrade of LHCb is purely a question of the detector being
able to profit from a higher peak luminosity. 
Section~\ref{sect:lumi} will describe the conditions as a function of
the delivered peak luminosity, and the limitations of LHCb to efficiently 
exploit an increase in luminosity. 

The baseline upgrade scenario of the detector to SuperLHCb will be
discussed in section~\ref{sect:super}, followed by expectations of
yields for some selected physics channels in comparison with the proposed
SuperKEKB performance in section~\ref{sect:yield}. The conclusions will be 
presented in section~\ref{sect:conclusions}.

\section{Expected performance of LHCb}
The expected performance of LHCb is determined by generating pp interactions
using the PYTHIA 6.2 generator~\cite{pythia}, with the predefined option 
MSEL=2. To extrapolate to 14 TeV CM the value of the $p_{\rm T}^{min}$ 
parameter has been tuned as a function of energy to existing data~\cite{ptmin}.
The resulting charged track multiplicities in the acceptance of the 
spectrometer are $\sim25\%$ larger than a similar tuning of CDF~\cite{cdf}.
The particles are propagated through a detailed detector description using 
GEANT. 
Pileup in a bunch crossing, and spill-over from preceding and following 
bunches is included. Trigger studies have shown that the events written to
storage are dominated by $\rm b\bar{b}$-events, hence all background is assumed
to originate from $\rm b\bar{b}$-events, of which the equivalent of
about 13 minutes of LHCb running have been fully simulated.

\subsection{BR(B$_s\rightarrow \mu^+\mu^-)$}
\begin{figure}[h]
\centering
\includegraphics[width=80mm,clip=]{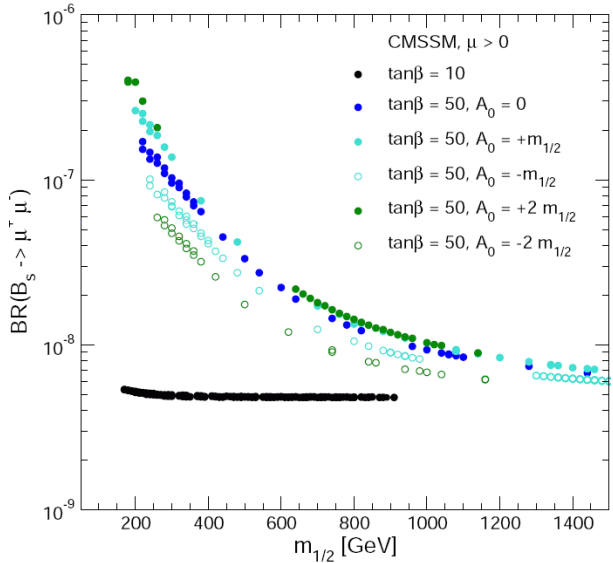}
\caption{The CMSSM prediction for BR(B$_s\rightarrow \mu^+\mu^-)$ as a function of the gaugino mass $m_{1/2}$ from~\cite{ellis}.} 
\label{ellis}
\end{figure}

\begin{figure}[h]
\centering
\includegraphics[width=80mm]{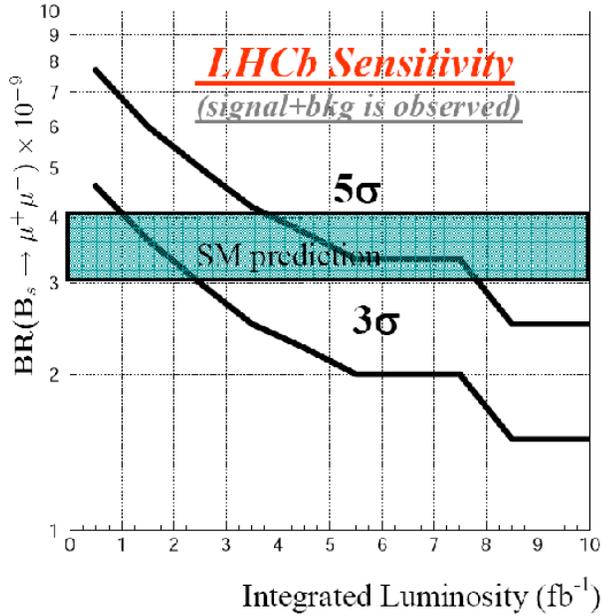}
\caption{The LHCb reach to observe ($3\sigma$) or discover ($5\sigma$)
the  BR(B$_s\rightarrow \mu^+\mu^-)$ as a function of integrated luminosity.}
\label{mumu}
\end{figure}

The rare loop decay of B$_s\rightarrow \mu^+\mu^-$ is sensitive to extensions
of the Standard Model (SM) through loop corrections. Within the SM the 
decay rate has been computed~\cite{brmumu} to be 
BR(B$_s\rightarrow \mu^+\mu^-)=(3.4\pm 0.4)10^{-9}$. NP physics beyond the SM
can increase this BR. In the  minimal super-symmetric extension 
of the SM (MSSM) 
the BR increases as $\rm tan^6\beta$, where $\rm tan\beta$ is the ratio of the 
Higgs vacuum expectation values. Hence, this makes the BR sensitive to 
models which prefer a relatively large $\rm tan\beta$. As an example 
figure~\ref{ellis} shows the expected BR as a function of the gaugino
mass in the framework of a constrained minimal super-symmetric extension 
of the SM (CMSSM)\cite{ellis}. 
The experimental challenge lies in the rejection of background, which
is predominantly due to two muons which combine to form a good vertex with
a signal mass. The muons originate either from semi-leptonic B-decays, or 
are due to misidentification of hadrons. LHCb combines a good
invariant mass resolution, $\sigma(\rm M_{\mu\mu})\approx 20$ MeV, and
excellent vertex resolution. In addition, the trigger can accept events
with $p_{\rm T}^{\mu}\geq 1$ GeV. Figure~\ref{mumu} shows the sensitivity~\cite{mumupap}
of BR(B$_s\rightarrow \mu^+\mu^-$) as a function of integrated luminosity.
Within the first years of running LHCb should be able to probe the whole
CMSSM parameter space for large $\rm tan\beta$ values via this rare loop decay.

\subsection{NP effects in $\rm B\rightarrow K^*(K\pi)\mu^+\mu^-$}
\begin{figure}[h]
\centering
\includegraphics[width=80mm,clip=]{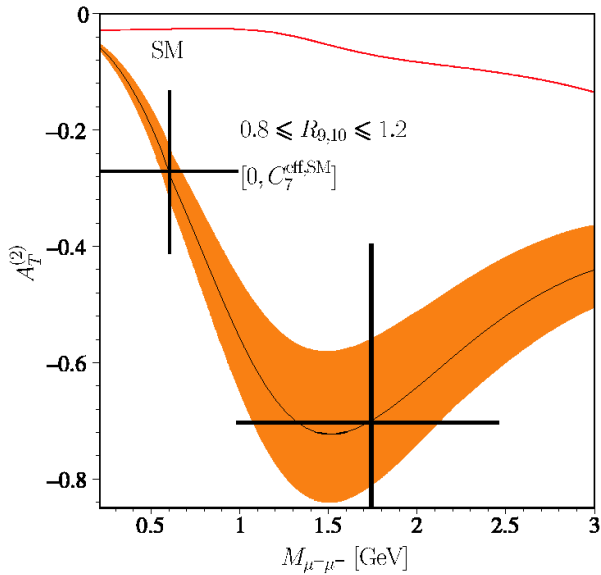}
\caption{$\rm A^{(2)}_T$ as a function of the dimuon mass for the SM 
(top curve), and 
in the presence of NP contributions to the Wilson coefficients $C_7,~C_9$ and 
$C_{10}$ as described in~\cite{matias}. The data points indicate the 
expected 95$\%$ confidense level sensitivity of LHCb for an integrated 
luminosity of $\rm 10~fb^{-1}$.} \label{f0502060}
\end{figure}

While it is shown in the previous section that LHCb is very sensitive to NP
effects at large $\rm tan\beta$ with a modest
integrated luminosity, this section will explore the sensitivity to small
$\rm tan\beta$ parameter space using the second transversity amplitude 
$\rm A^{(2)}_T$\cite{matias} in the decay $\rm B\rightarrow K^*(K\pi)\mu^+\mu^-$.
Figure~\ref{f0502060} shows $\rm A^{(2)}_T$ as a function of the dimuon mass for
both the SM expectation, and for a representative choice of NP parameters,
notably $\rm tan\beta=5$, 
which do take into account the constraints from present observations. 
Note that the whole region between the shown NP curves and the SM are filled by
solutions consistent with the constraints.
The expected LHCb 95$\%$ confidence interval sensitivity for $10\rm~fb^{-1}$ 
has been superimposed~\cite{ulrik}, assuming our measurements will fall precisely 
on the chosen NP expectation. 
While $\rm 10~fb^{-1}$ might allow to observe a hint of NP, 
an ten fold increase in statistic will allow a real observation of NP
if nature has chosen this particular constellation.

\subsection{NP in $b\rightarrow s\bar{s}s$ transitions}
Arguably the most intriguing hint of NP contributions to virtual loops
in B-decays comes from the discrepancy between 
$\rm sin 2\beta$ measured in the time dependent CP asymmetries in $b\rightarrow c\bar{c}s$ 
and in $b\rightarrow s\bar{s}s$ transitions. The later cannot decay via
a tree diagram in the SM, and hence is sensitive to NP contributions in its 
loop decay diagrams. The HFAG~\cite{hfag} averages for 
$\rm sin 2\beta(B\rightarrow J/\psi K^0_S)=0.668 \pm 0.026 $, and 
$\rm sin 2\beta^{eff}(B\rightarrow \phi K^0_S)=0.39 \pm 0.18 $. Although
the discrepancy is not statistically significant, all $b\rightarrow s\bar{s}s$
show a value of $\rm sin 2\beta^{eff}$ which is lower than the tree counterpart.
The expected sensitivity of LHCb for $\rm 10~fb^{-1}$ is
$\sigma(\rm sin 2\beta^{eff}(B\rightarrow \phi K^0_S))=\pm0.14$, while 
B-factories for a combined integrated luminosity of $2~\rm ab^{-1}$ expect
an error of $\pm 0.12$ in $\rm sin 2\beta^{eff}(B\rightarrow \phi K^0_S)$.

In addition LHCb has access to measuring the time dependent CP  asymmetries in
$\rm B_s$-decays, which give access to the CP violating weak phase $\phi$.
While $\phi_d^{\rm SM}({\rm B}\rightarrow J/\psi\rm K^0_S)=2\beta$,
 $\phi_s^{\rm SM}({\rm B}\rightarrow J/\psi\phi)=2\chi$, which is
constrained to -0.035$\pm^{0.014}_{0.006}$ by a fit to the unitary triangle
within the SM\cite{ckmfitter}. NP in the $\rm B_s\leftrightarrow \bar{B}_s$ 
mixing box diagram could enhance $\phi_s$. With a modest integrated luminosity
of $0.5~\rm fb^{-1}$ LHCb is expected to reach a sensitivity of
$\sigma(\phi_s(\rm{B_s}\rightarrow J/\psi\phi))=0.046$~\cite{psiphi}. Already this sensitivity
will constrain the parameters space of many extensions of the SM~\cite{ligeti}.
The golden hadronic 
counterpart is the decay $\rm B_s \rightarrow\phi\phi$, which can only proceed
via loop diagrams in the SM. In addition there is a cancellation of the
$\rm B_s$ mixing and decay phase in the SM~\cite{raidal}, which makes that
$\phi_s(\rm B_s\rightarrow \phi\phi)\approx 0.$ 
The BR$({\rm B_s}\rightarrow J/\psi\rm (\mu^+\mu^-)\phi(K^+K^-))$ is a factor eight larger than 
BR(${\rm B_s}\rightarrow \rm \phi(K^+K^-)\phi(K^+K^-))$. In addition, 
as will be explained in the next section, this
channel is much harder to trigger efficiently than channels with 
muons in the final state.
As a consequence, LHCb expects $\sigma(\phi_s^{\phi\phi})=0.054$~\cite{phiphi}
for an integrated luminosity of $\rm 10~fb^{-1}$. 
Even a factor of twenty increase in statistics will result in an 
experimental error on $\phi_s^{\phi\phi}$ which is larger than the theoretical 
error.

\section{The Luminosity Upgrade}
\label{sect:lumi}
\begin{figure}[h]
\centering
\includegraphics[width=80mm,clip=]{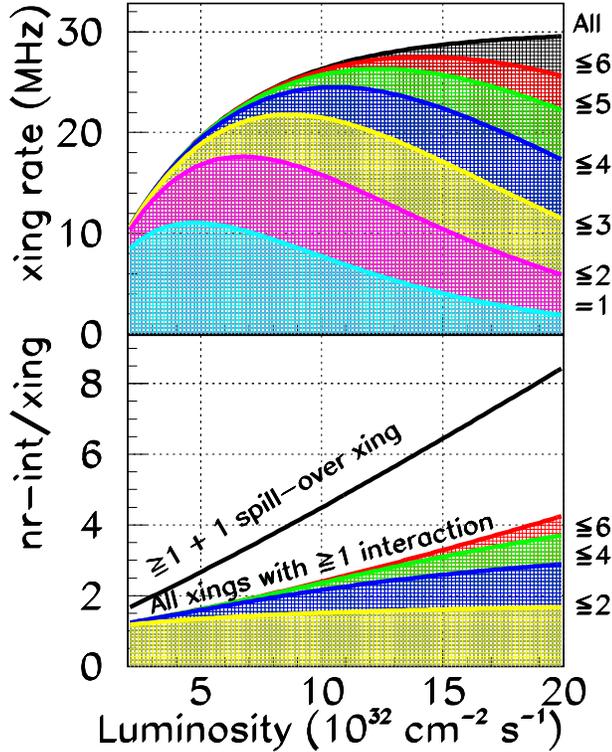}
\caption{Top plot shows the number of crossings with visible pp-interactions as
a function of luminosity. The bottom plot shows the average number of visible
pp-interactions per crossing, for events with at least one pp-interaction.
} \label{poisson}
\end{figure}

Before going into details about what is limiting the LHCb detector to
already profit from day one from larger luminosities, what follows 
is a brief description of the experimental environment at the LHCb
interaction point as a function of luminosity.
As already mentioned in the introduction, the LHC machine has been designed
to deliver a luminosity up to $10^{34}\rm cm^{-2}s^{-1}$ at a
General Purpose Detector (GPD). The optics
around the LHCb interaction point (P8) allows LHCb to run at a luminosity
up to $50\%$ of the luminosity available at a GPD. Hence, the nominal
LHC machine could deliver luminosities up to $5.10^{33}\rm cm^{-2}s^{-1}$
at P8\footnote{
While the nominal LHC is sufficient for the LHCb upgrade, there is
a proposal to increase the nominal luminosity of the machine to 
$8.10^{34}\rm cm^{-2}s^{-1}$, the SLHC, around the middle of the next decade.
The bunch separation for LHCb will remain 25 ns, but there are two 
schemes to fill the bunches. The preferred scheme will use large currents
in the even bunches, and low current in the odd bunches. Since P8
is displaced relative to the GPD interaction points by 1.5 bunch spacings,
it will result in colliding odd with even bunches in P8. In the GPD collision
points the collisions are odd$\times$odd and even$\times$even. This will
allow LHCb to choose its luminosity using the current in the odd bunches.
A GPD will ignore the odd$\times$odd interactions, since it will contribute
a luminosity at least a factor 400 smaller than what is obtained in  
the even$\times$even collisions. 
}. 
The bunch crossing rate at P8 is given by the LHC machine to be 40.08 MHz,
while 2622 out of the theoretically possible 3564 crossings~\cite{jorgen}
have protons in both bunches. 
Hence, the
maximum rate of crossings with at least one pp interaction is $\sim 30$ MHz.
The expected inelastic pp cross-section is 79.2 mb, of which 63 mb has at
least two charged particles which can be reconstructed, the so-called
visible cross-section. Figure~\ref{poisson} show the number of crossings
with at least one visible interaction and
the mean number of visible interaction per crossing
as a function of luminosity. 
Note that increasing the luminosity from
$(2\rightarrow 10).10^{32}\rm cm^{-2}s^{-1}$ will only increase
the mean number of interactions per crossing by a factor two, since
the number of crossings with at least one interactions increases from 
$10\rightarrow 26$ MHz. While the increase in occupancy for
detectors which are only sensitive to pileup is minimal, spill-over
increases linearly with luminosity as is indicated in the bottom plot of 
figure~\ref{poisson}.

\subsection{The LHCb Trigger}
\begin{figure}[h]
\centering
\includegraphics[width=80mm,clip=]{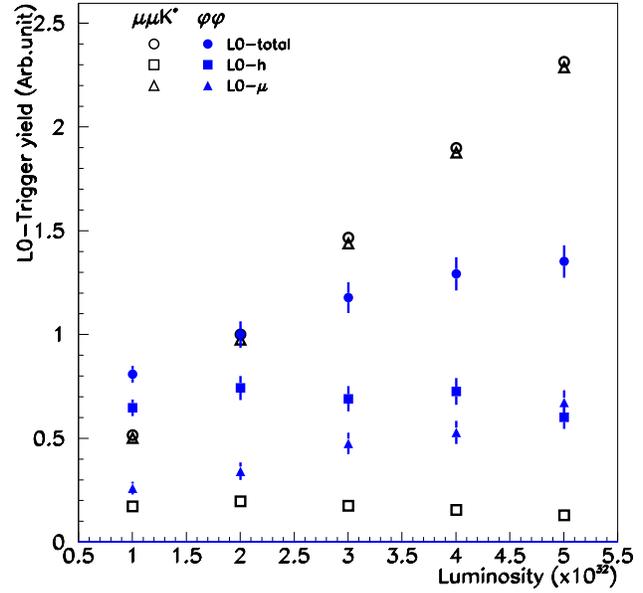}
\caption{The L0-trigger yield as a function of luminosity for two 
decay channels: $\mu\mu\rm K^*$ (open points) and $\phi\phi$ (closed points).
The total L0-trigger yield, and the contributions from the L0-hadron and muon
triggers are shown separately.
} \label{l0acc}
\end{figure}

LHCb has a two level trigger system, called Level-0 (L0) and the High Level
Trigger (HLT). L0 is a trigger implemented in hardware, and its purpose is 
to reduce the rate of crossings with interactions to below a rate of 1.1 MHz.
This is the maximum rate at which all LHCb data can be readout by the
front-end (FE) electronics.
L0 reconstructs the highest $E_{\rm T}$ hadron, electron and
photon, and the two highest $p_{\rm T}$ muons. It triggers on events with
a threshold of typically $E_{\rm T}^{\rm hadron}\gtrsim 3.5$ GeV, 
$E_{\rm T}^{\rm e,\gamma}\gtrsim 2.5$ GeV, and $p_{\rm T}^{\mu}\gtrsim 1$ GeV
at $2.10^{32}\rm~ cm^{-2}s^{-1}$.
Figure~\ref{l0acc} shows the yield of L0-triggered events, normalized to 
their yield at $2.10^{32}\rm~ cm^{-2}s^{-1}$ as a function of the luminosity for a
leptonic and a hadronic B-decay channel.
The L0-hadron trigger absorbs $\sim 70\%$ of the L0 bandwidth at 
$2.10^{32}\rm~ cm^{-2}s^{-1}$, and its threshold is already larger than
half the B-mass. The increase in mainly the rate of visible pp interactions
requires an increase in the threshold, and the resulting loss in efficiency 
nullifies the increase in luminosity, resulting in an almost constant yield
for the hadron trigger. Contrary, the muon trigger absorbs only $\sim 15\%$ of
the available bandwidth at $2.10^{32}\rm~ cm^{-2}s^{-1}$, at which rate it already has a efficiency
around $90\%$ for leptonic B-decays. For larger luminosities the loss
in efficiency is minor, showing an almost linear dependence of its yield
on luminosity. Note that at a luminosity of $5.10^{32}\rm~ cm^{-2}s^{-1}$
about half the yield in $\rm B_s\rightarrow\phi\phi$ is due to the
muon trigger on the leptonic decay of the tagging B. 

After L0, all detectors are readout, and full event building is performed
on the CPU nodes of the Event Filter Farm (EFF). 
The HLT consists of a C++ application which is running
on every CPU of the EFF, which contains between 1000 and 2000 multi-core 
computing nodes.
Each HLT application has access to all data in one event, and thus in principle
could be executing the off-line selection algorithms, which would render
it a 100$\%$ trigger efficiency by definition. But given the 1 MHz 
output rate of L0 and the limited CPU power available, the
HLT aims at rejecting the bulk of the events by using only part of the
full information which is available. 
The HLT starts with so-called alleys, where 
each alley addresses one of the trigger types of the L0-trigger, enriching
the B-content of the events by refining the L0 objects, and adding
impact parameter information. 
If an event is selected by at least one alley,  
it is processed by the inclusive triggers, where specific resonances
are reconstructed and selected, and the exclusive triggers, which aim
to fully reconstruct B-hadron final states. 
Events will be written to storage with a rate of $\sim 2$ kHz. 

\begin{figure}[h]
\centering
\includegraphics[width=80mm,clip=]{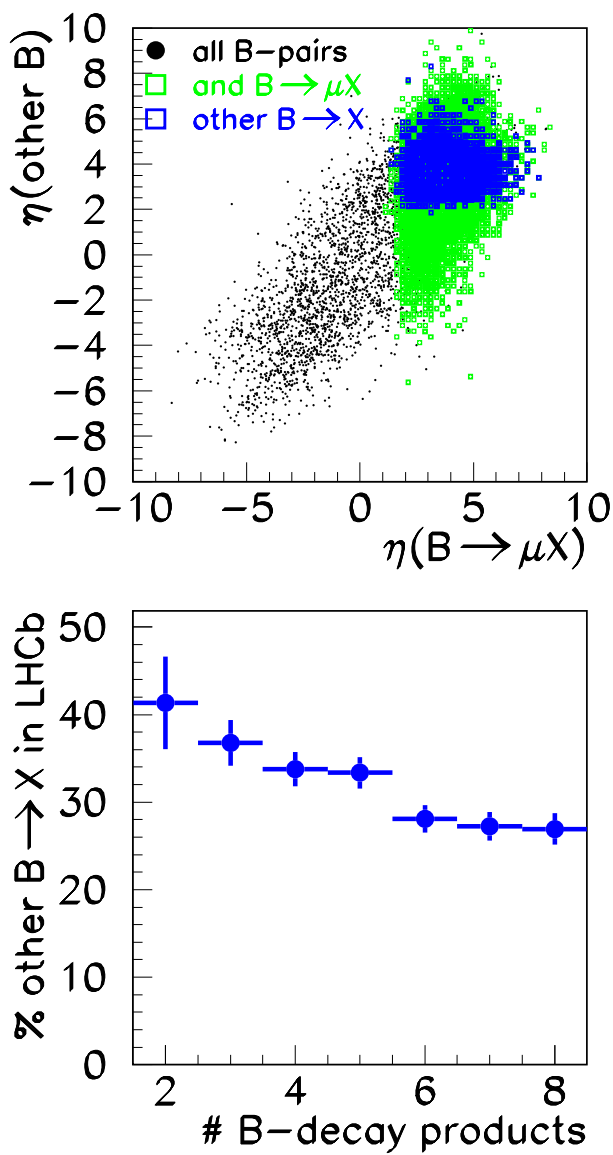}
\caption{Top plot shows the pseudo rapidity ($\eta$) correlation of a 
$\rm B\bar{B}$-pair. Of all the produced $\rm B\bar{B}$-pairs (black dots), the single
muon trigger selects the pairs of which one B has its decay-muon inside the
acceptance of LHCb (green squares). The blue squares indicate the rapidity
of the other B, if all its decay products are inside the LHCb acceptance.
The bottom plot shows the fraction of ``other'' B-decays in the LHCb acceptance
as a function of the number of B-decay products.
} \label{inclmu}
\end{figure}

As is shown in figure~\ref{l0acc}, even hadronic B-decays receive a 
considerable fraction of their L0 efficiency due to the muon trigger which
is usually fired by a leptonic decay of the opposite B. Hence, in the 2 kHz
output rate about half is reserved for events with a large $p_{\rm T}$ muon
with a significant impact parameter. 
Simulation shows that 
900 Hz of single muon triggers contain $\sim 550$ Hz of true 
$\rm B\rightarrow \mu X$ decays. 
Figure~\ref{inclmu} illustrates the
yield of B-decays, of which the decay products are 
fully contained  in the LHCb acceptance, while the event is triggered on the
semi-leptonic B-decay of the other B-hadron.
It shows the correlation of the $\rm B\bar{B}$-pair in pseudo rapidity, 
which makes that when the lepton from a semi-leptonic decay of one of
the B-mesons is in the acceptance of LHCb, there is a $30-40\%$ probability
that all the decay product of the opposite B are also contained in the spectrometer.
Hence, just the inclusive muon trigger will already provide a rate of  
$\sim 10^9/\rm fb^{-1}$ fully contained B-decays, with
a tagging performance of $\epsilon\rm D^2\approx 0.15$ due to the presence
of a large $p_{\rm T}$ muon. 

The typical efficiency of the whole trigger chain for hadronic, radiative and 
leptonic B-decays is 25-30$\%$, 30-40$\%$ and 60-70$\%$ respectively.
An upgrade of the trigger should not only be able to cope with
larger luminosities, but should be designed to at least gain a factor two
for hadronic B-decays like $\rm B_s\rightarrow\phi\phi$.

\subsection{Tracking and Particle Identification}
The tracking of LHCb commences by reconstructing all tracks inside the VELO.
The VELO is based on silicon (Si)-sensors, and the channel occupancy at 
$\rm 2.10^{32}~\rm cm^{-2}s^{-1}$ is $\sim 1\%$, which is kept roughly constant
as a function of the radius due to the layout of the strips.
This occupancy increases to $\sim 3\%$ at $\rm 2.10^{33}~\rm cm^{-2}s^{-1}$.
As a result the tracking performance~\cite{matt} in the VELO looses only
$2.7\%$ in efficiency for this factor ten increase in luminosity, while
using reconstruction code tuned for the low luminosities. This is as 
expected from figure~\ref{poisson}, since the VELO electronics has a limited
sensitivity to spill-over, and as a consequence only $27\%$ of its
hits at $\rm 2.10^{33}~\rm cm^{-2}s^{-1}$ are due to spill-over.

To assign momentum to the VELO tracks, every track is combined with a hit
in the T-stations, located behind the magnet, in turn. Around the trajectory 
defined by the VELO track and a T-hit, a search is performed in the other 
tracking stations including TT. In addition there is also a stand-alone
pattern recognition performed in the T-stations, mainly to recover 
$\rm K^0_S$ which decay behind the VELO. The outer part of the T-stations (OT) are 
constructed
of 5 mm diameter straws, with a drift-time up to 50 ns. 
Including the effect of the length of the 2.4 m long wires, 
this requires a readout gate of three 
consecutive crossings to obtain the maximum efficiency. As a consequence,
the OT occupancy rises from $6\rightarrow 25\%$ for a ten-fold luminosity increase
from $\rm (2\rightarrow 20).10^{32}~\rm cm^{-2}s^{-1}$. 
At $\rm 2.10^{33}~\rm cm^{-2}s^{-1}$  $60\%$ of the OT hits are due to spill-over.
Both TT and the inner part of the T-stations 
(IT) are made of Si-sensors. At $\rm 2.10^{33}~\rm cm^{-2}s^{-1}$  $44 (25)\%$ of the TT(IT) hits are due to spill-over.

\begin{figure}[h]
\centering
\includegraphics[width=80mm,clip=]{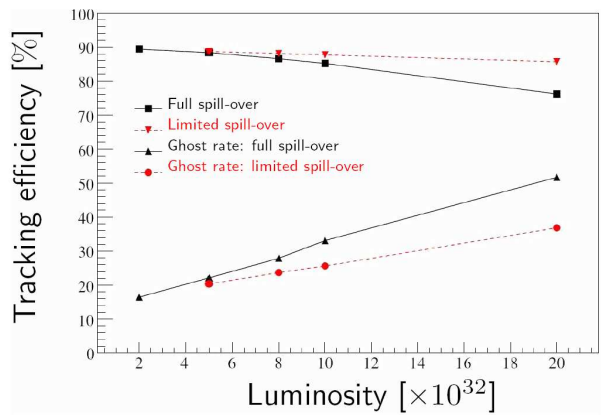}
\caption{The tracking efficiency as a function of luminosity. Limited 
spill-over uses the same spill-over as obtained at a luminosity of 
$\rm 2.10^{32}~\rm cm^{-2}s^{-1}$, irrespective of the luminosity.
} \label{teff}
\end{figure}

The tracking performance as a function of luminosity is shown
in figure~\ref{teff}.
Above a luminosity of $5.10^{32}~\rm cm^{-2}s^{-1}$ the difference between
the tracking performance with full spill-over and a spill-over equivalent to
the spill-over at $\rm 2.10^{32}~\rm cm^{-2}s^{-1}$ is clearly visible.
The per track loss in tracking efficiency from 
$\rm (2\rightarrow 10).10^{32}~\rm cm^{-2}s^{-1}$ is $\sim 5\%$, which is 
a small price to be paid,
even for a 4-prong decay, compared to the factor 5 increase in luminosity.
However, an additional increase of a factor two in luminosity would result
in a loss of 36$\%$ of the 4-prong decays, hence would almost eliminate the
additional factor two increase. 

The electronics of the RICH detector is virtually insensitive to spill-over.
The increase in occupancy for an event with a $\rm B\bar{B}$-pair and its
pileup events is only a factor 2.5 for the factor ten increase in luminosity
to $\rm 2.10^{33}~\rm cm^{-2}s^{-1}$. This is even a smaller increase than
shown in figure~\ref{poisson}, since pp-interactions producing a 
$\rm B\bar{B}$-pair cause about twice the occupancy compared to visible
pp-interactions. The efficiency to positively identify a kaon degrades
by $10\%$ for the 10 fold increase in luminosity. The loss is dominated
by the degradation of the RICH1 performance, which has the higher
backgrounds and occupancies. In the above simulation the effect of 
inefficiency due to overflowing buffers at high occupancy has not been
taken into account, since it will be argued in the next section that the 
front-end
electronics will have to be replaced to allow the trigger to be upgraded.

The muon chambers and the calorimeters both have negligible sensitivity
to spill-over, and hence the increase in occupancy follows the same trend
as that of the RICH. Their performances development as a function of luminosity
is more related to dead time inefficiency and radiation damage, rather than
occupancy per bunch crossing. These effects are not simulated in the MC, and
hence no reliable performance as a function of luminosity is available.

\section{The SuperLHCb Detector}
\label{sect:super}
In the previous section it was shown that the sub-system of LHCb which
does not scale in performance with an increased luminosity is the trigger, and
in particular the trigger for hadronic B-decays which will not be able to
retain its efficiency for larger luminosities.
Since the trigger efficiency for hadronic B-decays is expected to be
$25-30\%$, the goal of an upgrade of the trigger should also be to to improve
on the hadron trigger efficiency by at least a factor two. 
At 14 TeV center of mass pp collisions $\sigma_{b\bar{b}}$ is assumed to be 
500 $\mu$b. Hence, with a luminosity of $\rm 2.10^{33}~\rm cm^{-2}s^{-1}$
there will be $10^6$ $b\bar{b}$-pairs produced in the LHCb interaction point
per second, of which $43\%$ will have at least one B-hadron with a polar
angle below 400 mrad, i.e. pointing in the direction of the spectrometer.
Hence, an efficient and selective trigger should already at a very large rate
be able to distinguish between wanted and unwanted B-decays. 
Pilot studies on improving the trigger all show that the only way
to be able to provide adequate selectivity of the trigger, and maintain
large efficiency for hadronic B-decays is to be able to measure both
the momentum and impact parameter of B-decay products simultaneously.

The present FE-architecture imposes that the detectors which do not
participate in the L0-trigger can only be read-out with a  maximum event rate of 
1.1 MHz, and that the L0-latency available for making the L0 decision, which
is now 1.5$\mu$s, can be stretched to a few $\mu$s at most.
The algorithms required to efficiently
select B-decays require latencies far superior to what is available with
the present architecture. 

Hence, SuperLHCb has opted for a FE-architecture which
requires all sub-detectors to read-out their data at the full 40 MHz rate of
the LHC machine. The data should be transmitted over optical fibers to
an interface board (TELL40~\cite{guido}) between the FE and a large EFF. The trigger
algorithm is then executed on the EFF, much like the present HLT.
Technology tracking estimates show that by 2013 SuperLHCb should be able to
acquire sufficient CPU power to be able to perform a HLT like trigger
on a large CPU farm. However, in case the EFF at the start of SuperLHCb would 
be undersized, the TELL40 boards will be equipped with a throttle to
prevent buffer overflows. This throttle should also include an event selection
much like the present L0, based on the data available in the TELL40 boards, to
enrich rather than just pre-scale events. At a luminosity of 
$\rm 6.10^{32}~\rm cm^{-2}s^{-1}$
and an assumed CPU power able to process 5 MHz of events, the
trigger efficiency is $66\%$ for the channel 
$\rm B_s\rightarrow D_s^\mp K^\pm$. For this simulation the throttle requires  
at least one HCAL-cluster with $E_{\rm T}^{\rm hadron}> 3$ GeV, 
which has an efficiency of $76\%$ for this signal. 
Note that a 2 GeV requirement
would correspond to 10 MHz of input rate into the EFF, while it would
increase the efficiency from $76\rightarrow 95\%$ at the start of the EFF, 
while with LHCb running at $\rm 2.10^{32}~\rm cm^{-2}s^{-1}$
the equivalent efficiency for this channel at the start of the HLT is 
only $39\%$.

The upgraded FE-architecture requires that the FE-electronics of
all sub-detectors
needs to be replaced,
with the exception of the muon chambers which already have the 40 MHz capability.
The Si-detectors, which cover the areas close to the beam, will
suffer from a five fold increase in allowed radiation dose, 
and hence need to be replaced by more radiation resistant technologies. 
For the RICH the photon detection and the FE-electronics is combined
in a Hybrid Photo Detector, which needs to be replaced entirely.
The OT requires the replacement of its FE-boards. Running these detectors with
a slightly faster gas, combined with taking advantage of being able to
pre-process spill-over in the TELL40 boards could
reduce the occupancy from $25\rightarrow 17\%$ at 
$\rm 2.10^{33}~\rm cm^{-2}s^{-1}$. 
This could be combined with enlarging the coverage of IT, to reduce
the occupancy close to the beam even further. 
The M1 muon chamber, which is located just before the Calorimeter, would
suffer from a too high occupancy, and will be removed. It now serves
to provide an improved momentum measurement in L0, which will no longer
be necessary. 
The resolution in the inner part of the Calorimeter will degrade with radiation.
It will have to be replaced with a more radiation tolerant technology.
This might also allow LHCb to extend the calorimeter coverage down to
$\eta=5$ from the present maximum pseudo rapidity of 4.2, i.e. an increase
in coverage of 25$\%$.
Last but not least, all results presented from simulation studies did not
attempt to adapt the algorithms to a higher occupancy environment, hence
they are considered to be conservative.

\section{Projected Yields of SuperLHCb}
\label{sect:yield}
\begin{figure*}[t]
\centering
\includegraphics[width=135mm,clip=, angle=-90.]{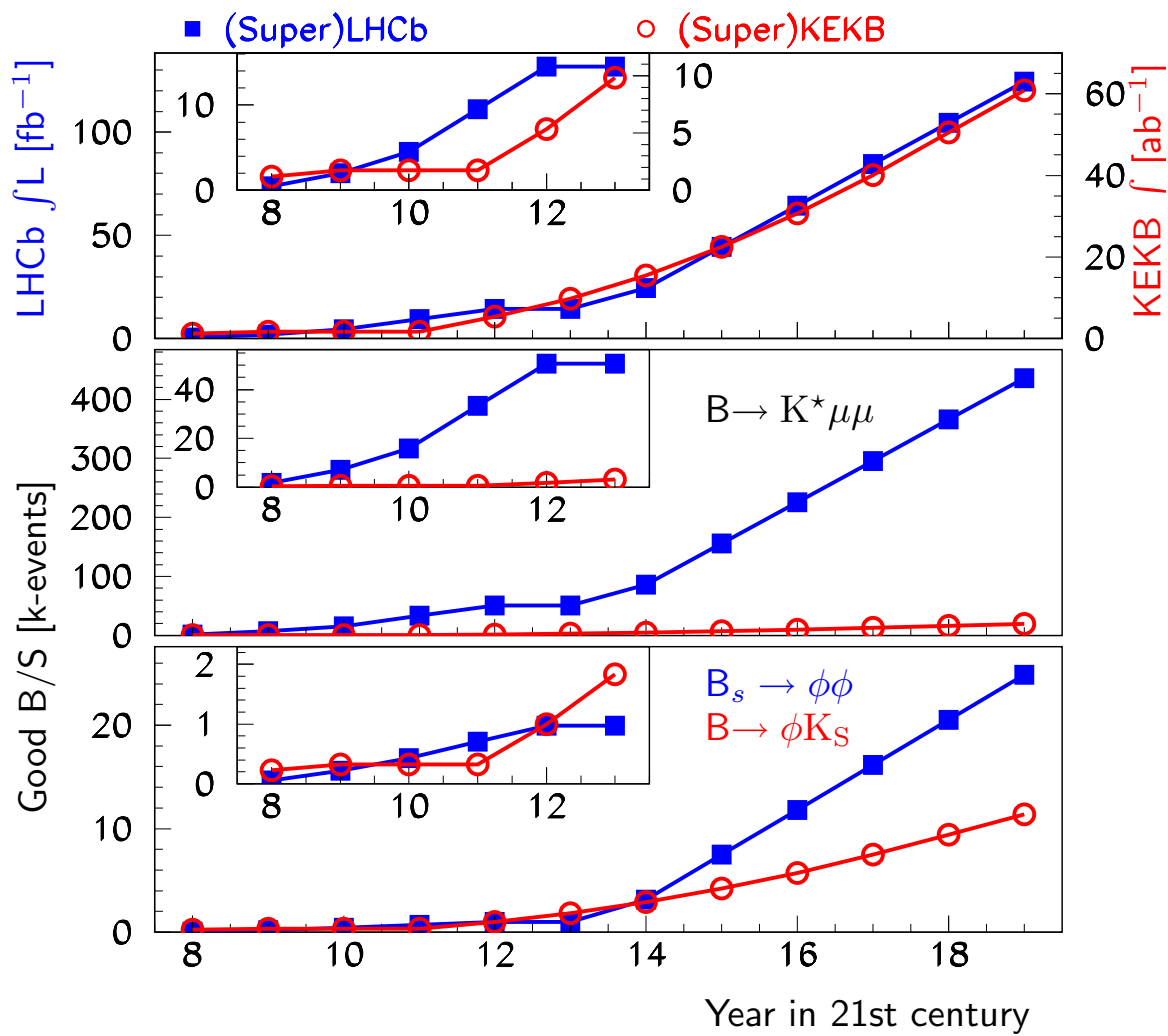}
\caption{Top plot shows the projected integrated luminosity for LHCb and
Super LHCb, compared to (Super)KEKB. The middle and bottom plots shows 
the expected yield for the channels indicated, after trigger and 
strict off-line selection to obtain a good Background/Signal ratio. 
The yield of $\rm Bs\rightarrow\phi\phi$ and $\rm B\rightarrow\phi K_S$ has
been multiplied by 
$\epsilon D^2$, 0.07 and 0.3 for LHCb and KEKB respectively, 
to take into account the better tagging performance at a B-factory.
} \label{yield}
\end{figure*}

The LHC machine schedule assumes that first collisions at 14 TeV will be
delivered in the summer of 2008. The top plot of figure~\ref{yield} shows 
the expected integrated luminosity profile, which assumes that LHCb will
run at $\rm (2-5).10^{32}~\rm cm^{-2}s^{-1}$, and that the machine and experiment
will only slowly ramp up to the full capability in 2011. 

LHCb would then
run at maximum luminosity for two years, and then have a one year shutdown
to change over to the new FE-architecture in 2013. In 2014 it assumes to run at
half of its full capability, after which it accumulates $20~\rm fb^{-1}$ per 
year for the rest of the next decade. 

For comparison the running scenario
of the proposed SuperKEKB~\cite{skekb} is shown in the same plot.
The closed squares (open circles) show the information of LHCb (KEKB).
The middle and bottom plots show
the expected yield for $\rm B\rightarrow K^*\mu\mu$, $\rm Bs\rightarrow\phi\phi$ and $\rm B\rightarrow\phi K_S$\, after trigger and 
strict off-line selection to obtain a good Background/Signal ratio. 
For the channels which require tagging to perform the time dependent CP asymmetry analysis, the yield has 
been multiplied by the effective tagging efficiency 
$\epsilon D^2$, 0.07 and 0.3 for LHCb and KEKB respectively, 
to take into account the better tagging performance at a B-factory.

\section{Conclusions}
\label{sect:conclusions}
The last few years have seen an impressive progress in precision measurements
of B-decays, notably from the B-factories and the Fermilab collider.
The remarkable agreement between CP conserving and violating observables
indicates that the main source 
of CP-violation can be attributed to the KM-mechanism~\cite{km}.
In their quest for discovering NP, the new generation of experiments have
to be able to detect small deviations from the SM, which requires increasingly
larger data sets. LHCb is nearing the end of its construction, and will
be ready to look for NP with a projected integrated luminosity of
around $10~\rm fb^{-1}$ in the years to come. This paper describes the
way LHCb can be upgraded to be able to have access to NP even beyond the
possibilities of the first phase of LHCb. 

The main component of LHCb which limits it to profit from the available
nominal luminosity of the LHC machine is the hadron-trigger. Consequently
SuperLHCb will have a new FE-architecture and trigger which aims at
being able to cope with luminosities around $\rm 2.10^{33}~\rm cm^{-2}s^{-1}$,
and which will have a hadron trigger efficiency twice larger than the
present trigger, resulting in a twenty fold increase in hadronic B-decays
available for analysis. In addition, the leptonic decay channels will profit
from an increase in luminosity at least linearly.

\begin{acknowledgments}
This paper would not have been possible without the work done by many of
my colleagues in LHCb, notably those who contributed to the ``1$^{st}$ LHCb
Collaboration Upgrade Workshop''~\cite{edinburgh}. I would like to thank
them all. I would like to thank Franz Muheim for carefully reading the 
manuscript.

\end{acknowledgments}

\bigskip 


\end{document}